\documentclass[apjl]{emulateapj}

\usepackage{hyperref}
\usepackage{amsmath}
\usepackage{amssymb}
\usepackage{graphicx}
\usepackage{color}
\usepackage{natbib}
\usepackage[normalem]{ulem}
\usepackage{multibib}

\newcommand{\Msun}{\mathrm{M}_\odot}

\newcommand{\MBH}{M_\mathrm{BH}}
\newcommand{\MNS}{M_\mathrm{NS}}
\newcommand{\chieff}{\chi_\mathrm{eff}}
\newcommand{\chiz}{\chi_\mathrm{BH,z}}

\def\be{\begin{equation}}
\def\ee{\end{equation}}
\def\ba{\begin{eqnarray}}
\def\ea{\end{eqnarray}}

\begin{document}

\title{GW200115: a non-spinning black hole -- neutron star merger}
\author{Ilya Mandel\altaffilmark{1,2,3}}
\email{ilya.mandel@monash.edu}
\author{Rory J.~E.~Smith\altaffilmark{1,2}}
\email{rory.smith@monash.edu}
\affil{$^1$Monash Centre for Astrophysics, School of Physics and Astronomy, Monash University, Clayton, Victoria 3800, Australia}
\affil{$^2$The ARC Center of Excellence for Gravitational Wave Discovery -- OzGrav}
\affil{$^3$Institute of Gravitational Wave Astronomy and School of Physics and Astronomy, University of Birmingham, Birmingham, B15 2TT, United Kingdom}

\begin{abstract}
GW200115 was the second merger of a black hole and a neutron star confidently detected through gravitational waves.  Inference on the signal  allows for a large black hole spin misaligned  with the orbital angular momentum, but shows little support for aligned spin values.   We show that this is a natural consequence of measuring the parameters of a black hole -- neutron star binary with non-spinning components while assuming the priors used in the LIGO-Virgo-KAGRA analysis.  We suggest that, a priori, a non-spinning binary is more consistent with current astrophysical understanding.
\end{abstract}

\maketitle

\section{Introduction}

Several black hole (BH) -- neutron star (NS) merger candidates were observed in the gravitational-wave (GW) data from the third science run of advanced LIGO and Virgo detectors.  These included events such as GW190814, in which the mass measurements of the lower-mass companion could be either an NS or a BH \citep{GW190814}, as well as lower-significance candidates with uncertain measurements, particularly GW190426\_152155 and GW190917\_114630 \citep{Abbott:2021-GWTC-2-1}.  Two events stood out as confident BH -- NS merger detections: GW200105 and GW200115  \citep{GW200105}.

The GW signature GW200105---a merger between a $\approx 9 \Msun$ BH and a $\approx 2 \Msun$ NS---constrained the BH dimensionless spin magnitude to low values ($\chi_\mathrm{BH}<0.3$ at 95\% confidence).  The effective spin, defined as 
\begin{equation}\label{eq:chieff}
\chieff \equiv \frac{(\MBH \vec{\chi}_\mathrm{BH} + \MNS \vec{\chi}_\mathrm{NS}) \cdot \hat{L}}{\MBH+\MNS},
\end{equation}
where $M_\mathrm{BH}$ and $M_\mathrm{NS}$ are the BH and NS masses, $\chi_\mathrm{BH}$ and $\chi_\mathrm{NS}$ are the corresponding dimensionless spins ($0 \leq \chi \equiv |\vec{\chi}| \leq 1$) and $\hat{L}$ is the unit vector along the orbital angular momentum, was centered on 0 for GW200105.  

In contrast, inference on GW200115---a merger between a $\approx 6 \Msun$ BH and a $\approx 1.4 \Msun$ NS---allowed for a much larger BH spin, with a median of 0.3 and a 90\% credible interval extending above 0.8.  The effective spin posterior encompassed zero, but centered on negative values, with a 90\% credible interval spanning $\chieff \in [-0.5,0.03]$.  The inferred preference for a BH spin anti-aligned with the orbital angular momentum (probability of 88\%) is surprising on astrophysical grounds \citep[e.g.,][]{Kalogera:2000}, and has led a number of authors to investigate the astrophysical implications and observational consequences of this apparent misalignment.

\citet{Fragione:2021} find that NS kicks following a Maxwellian distribution with a one-dimensional root mean square speed exceeding 150 km s$^{-1}$ are necessary for a non-negligible misalignment probability, but that in order to preserve the binary from disruption with such large kicks, the common envelope phase must be very efficient in hardening the binary.  \citet{Gompertz:2021} focus on the asymmetric natal kick accompanying the birth of the BH and find that in order for the BH to be significantly misaligned, the BH must have experienced a large natal kick, perhaps of hundreds of km s$^{-1}$.  \citet{Zhu:2021} focuses on the natal kick of the NS and finds that it had to be even larger  ($\sim 600$ km s$^{-1}$) if the BH spin and the orbit are misaligned by more than 90\,$\deg$.  Both \citet{Gompertz:2021} and \citet{Zhu:2021kn} conclude that the misalignment makes it less likely that the NS would be disrupted prior to plunging into the BH, making for an electromagnetically quiet merger.

Here, we determine that the support for large and negative BH spins is a natural consequence of analysing a BH -- NS merger with non-spinning components with the priors used in the LIGO-Virgo-KAGRA (LVK) analysis.  We analyse the correlation between spin and mass ratio in section \ref{sec:data} and show that the shape of the GW200115 posterior is expected for a BH -- NS merger with a non-spinning BH component and masses close to the maximum \textit{a posteriori} values of GW200115 when using \citet{GW200105} priors.  In Section \ref{sec:astro} we discuss the astrophysical context for non-spinning BH's in BH -- NS binaries, and propose alternative astrophysically motivated spin priors for the analysis of BH -- NS mergers.  Applying our spin priors constrains the black hole spin to be close to zero, consistent with expectations, and also leads to tighter constraints on the component masses, with $\MBH =7.0^{+0.4}_{-0.4}\,M_{\odot}$ (median and 90\% credible interval) and $M_{\text{NS}}=1.25^{+0.09}_{-0.07}\,M_{\odot}$,  which is typical of second-born neutron stars in Galactic double neutron star binaries \citep{Tauris:2017}.

\section{Data analysis}\label{sec:data}

There is a well-known correlation between the spin and mass ratio of binaries observed through gravitational waves from the inspiral phase of the coalescence \citep[e.g.,][]{CutlerFlanagan:1994,PoissonWill:1995,Baird:2013,Hannam:2013,Ng:2018}.  Although the mass ratio and spin-orbit coupling terms enter the waveform phase at different post-Newtonian order (1 pN and 1.5 pN, respectively), their contributions cannot be clearly distinguished for a signal with a limited signal-to-noise ratio.  In this section, we explain why this correlation naturally leads to an apparent preference for a misaligned solution for a source with negligible component spins.

\begin{figure}
\centering
\includegraphics[width=\columnwidth]{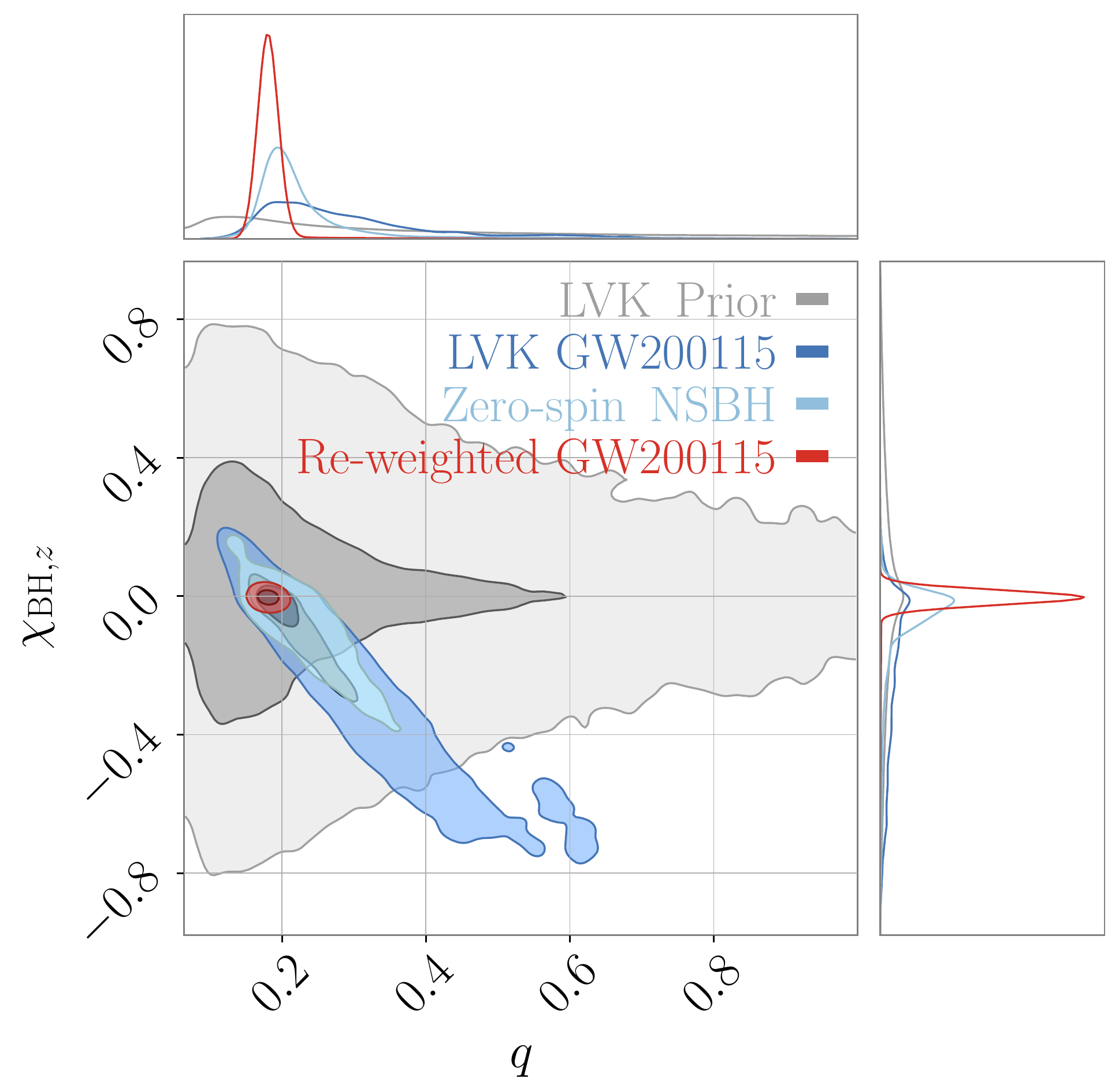}
\caption{Marginalized two- and one-dimensional prior (light/dark grey, for 86/39\% credible intervals, respectively) and posteriors in the space of mass ratio $q$ and BH spin projection along the orbital angular momentum $\chi_\mathrm{BH,z}$.  Dark blue shows the GW200115 posteriors as inferred in \citep{GW200105}, light blue shows those for a zero-noise mock signal with zero spins and similar component masses, and red shows the GW200115 posteriors re-weighted using the alternative priors proposed in section \ref{sec:astro}.}\label{figure:qchiz}
\end{figure}

We first show that the posterior obtained for GW200115 is similar to that obtained for an analysis of a mock BH -- NS system with zero component spin, and with component masses comparable to the maximum a posteriori values of GW200115. Figure \ref{figure:qchiz} displays the 2-dimensional and 1-dimensional posterior probability density functions on the the mass ratio $q\equiv \MNS / \MBH$ and the component of the black hole spin along the direction of the orbital angular momentum $\chiz \equiv \vec{\chi}_\mathrm{BH} \cdot \hat{L}$. Grey shows the priors from the \citet{GW200105, GWOSC} analysis, with the additional constraint placed on the component masses ($\MBH,\MNS>1\,M_{\odot}$) for consistency with our mock-data study. Dark blue shows the inferred posteriors for GW200115 using a combination of IMRPhenomXPHM and SEOBNRv4PHM \citep{Pratten:2021, Ossokine:2020} waveform models. In light blue, we show the posteriors for an injected waveform with non-spinning components of lab-frame mass $\MBH=6.94 \Msun$ and $\MNS=1.30 \Msun$, into a zero-noise realisation with detector noise power spectral densities identical to the ones used to analyze GW200115. All other signal parameters are the same as the maximum posterior values of GW200115 \citep{GWOSC}. The waveform model used for the injection and its recovery is IMRPhenomPv2 \citep{Hannam:2013waveform,Khan:2016}, which does not contain higher-order modes, unlike the IMRPheonmXPHM and SEOBNRv4PHM waveforms used in the original analysis. However, this is not expected to make a difference as there is no observable higher-mode content in the signal. In red, we show the posteriors for GW200115 obtained by reweighting the LVK samples \citep{GWOSC} with the alternative astrophysically motivated spin prior discussed in Sec~\ref{sec:astro}.

The GW200115 posteriors and those recovered for the mock injection are qualitatively similar.  Both peak at zero values of $\chiz$, the BH spin projected onto the orbital angular momentum.  Both posteriors show a clear anti-correlation between the mass ratio $q$ and $\chiz$.  Yet both posteriors are clearly asymmetric in $\chiz$, with support at negative but not positive values, despite the symmetric prior.

The anti-correlation between $q$ and $\chiz$ can be easily understood by considering the frequency-domain inspiral waveform $\tilde{h}(f;\vec{\theta}) = A(f) e^{i\psi(f;\vec{\theta})}$ in the stationary-phase approximation, where $\vec{\theta}$ denotes the signal parameters.  Two waveforms with different parameters $\vec{\theta}_1$ and $\vec{\theta}_2$ will have a high match if their phases $\psi(f)$ are nearly equal at frequencies where the detector has optimal sensitivity, the so-called bucket frequency.  This will, in turn, lead to small residuals between the waveforms and thus a high likelihood that a data set containing a signal with parameters $\vec{\theta}_1$ could be generated with model parameters $\vec{\theta}_2$.  

The post-Newtonian waveform phase can be expanded in a Taylor series around the bucket frequency $f_0$,
\begin{equation}\label{eq:Taylor}
\psi(f) = \psi(f_0) + \frac{d\psi}{df}\Big|_{f_0} (f-f_0) + \frac{d^2\psi}{df^2}\Big|_{f_0} \frac{(f-f_0)^2}{2} + ...
\end{equation}
The constant term $\psi(f_0)$ is ignorable when comparing two waveforms, since it can be absorbed into an overall phase offset.  The linear in frequency term effectively corresponds to a time offset and can be absorbed into the definition of the coalescence time.  Thus, the first relevant term is the quadratic one, and two waveforms will have a high match when their second derivatives are approximately equal in the bucket, 
\begin{equation}\label{eq:ddpsi}
\frac{d^2\psi(f;\vec{\theta}_1)}{df^2}\Big|_{f_0} \approx \frac{d^2\psi(f;\vec{\theta}_2)}{df^2}\Big|_{f_0}
\end{equation}
(see, e.g., section IV.B of \citealt{Psaltis:2020} for a longer discussion).  

We can thus quantify the correlation between $q$ and $\chiz$ by asking what $\chiz$ values would yield an accurate match to the signal from a binary with non-spinning components of fixed mass as we vary $q$.   The chirp mass $M_c \equiv \MBH^{3/5} \MNS^{3/5} (\MBH+\MNS)^{-1/5}$ is generally very accurately determined for low-mass GW events (the fractional 1-$\sigma$ uncertainty is $\sim 1\%$ for GW200115), so we assume that it is fixed (in practice, the small but non-negligible uncertainty in $M_c$ is partly manifest in the finite width of the $\chiz$--$q$ posterior perpendicular to the direction of correlation in figure \ref{figure:qchiz}).  We further assume that $\chi_\mathrm{NS} = 0$, since the NS spin is expected to be low in merging binaries (e.g., $\chi_\mathrm{NS} \lesssim 0.02$ in Galactic field double neutron stars, \citealt{KumarLandry:2019}) and its impact is, in any case, diluted by the smaller NS mass.  We consider a 1.5 order post-Newtonian expansion of $\psi(f)$ \citep{PoissonWill:1995}.  Under these assumptions, equation \ref{eq:ddpsi} determines the expected correlation between $\chiz$ and $q$ posteriors when a non-spinning BH merges with an NS:
\begin{eqnarray}
\chiz &=& \Big(19.1 (\eta_0^{-2/5}\eta^{3/5}-\eta^{1/5}) \nonumber \\
&+& 23.8 (\eta_0^{3/5} \eta^{3/5} - \eta^{6/5})\Big) \left(\frac{M_c}{\Msun}\right)^{-1/3}  \left(\frac{f_0}{100\, \mathrm{Hz}}\right)^{-1/3}  \nonumber\\
&-& 12.6 (\eta_0^{-3/5}\eta^{3/5}-1), \label{eq:chiz}
\end{eqnarray}
where $\eta\equiv q/(1+q)^2$ is the symmetric mass ratio and $\eta_0$ is its value for the presumed non-spinning signal.  While the exact slope of the correlation predicted by equation (\ref{eq:chiz}) does not perfectly match the slope of the blue posteriors in figure \ref{figure:qchiz}, the difference is consistent with a range of simplifications used here, including cutting off the waveform at the 1.5 post-Newtonian order.

Having analyzed the correlation between $\chiz$ and $q$, we now discuss the reasons why the $\chiz$ posterior is skewed toward negative values for an injected signal with zero component spins.  The analysis above already points to one such reason. As the mass ratio $q$ is decreased at fixed chirp mass, the total mass $M$ becomes large because  $M \equiv \MBH+\MNS = M_c (1+q)^{6/5} q^{-3/5}$. Differences between post-Newtonian orders scale as $(Mf)^{1/3}$, and so become more significant at a fixed bucket frequency. Thus, even if two waveforms have equal $d^2\psi/df^2$, differences in $d^3\psi/df^3$ are amplified at low $q$. Consequently, low $q$ (which correlates with large positive $\chiz$) models with matching $d^2\psi/df^2$ produce larger residuals and are disfavored when analyzing a signal from a binary with non-spinning components, whose posterior peaks near zero spin. 

The total mass of the binary impacts the system beyond the phasing or frequency evolution.  The total mass and spin set the maximum frequency reached at the end of the inspiral and the ringdown frequency \citep{Echeverria:1989}.  For low values of total mass, the frequency at the end of the inspiral, $\sim 4(\Msun/M)$ kHz, is too high to be directly observable by current detectors.  However, for low $q$, the total mass would increase at fixed $M_c$, bringing the signal termination frequency into the detectors' sensitive frequency band.  Not detecting these effects can therefore rule out low $q$ without discriminating between large $q$, explaining the asymmetry in the $q$--$\chiz$ posterior.

Finally, parameter constraints can lead to unanticipated priors. \citet{GW200105} describe the mass priors as uniform in component masses.  At first glance, this should correspond to a flat distribution on $q$.  However, there were additional priors cuts imposed: both masses were chosen to be between 0.5 and 22.95 $\Msun$ \citep{GW200105}, which leads to a prior distribution on $q$ which begins to drop off below $q \lesssim 0.1$.  Furthermore, there is an additional cut that $q\geq1/18$ to match the waveform family requirements.  In view of the $q$--$\chiz$ correlation, this reduction in the prior support at low $q$ disfavors high $\chiz$.
 
\section{Astrophysics}\label{sec:astro}

We argued above that the observed posterior on $\chiz$, with a peak near zero but asymmetric support at negative $\chiz$ values, is exactly what one should expect for the analysis of a merging BH -- NS binary with a non-spinning BH when using the LVK priors.  Now, we turn to the question of astrophysical expectations, which point strongly against a significant BH spin misaligned with the orbital angular momentum.

The main problem with a BH spin misaligned with the orbital angular momentum in a BH -- NS binary is not the misalignment, but the fact the BH is significantly spinning at all.  To see this, we consider the possible channels for BH -- NS formation (see \citealt{MandelFarmer:2018,Mapelli:2021} for reviews).  

A merging BH -- NS binary could form dynamically, through interactions with other stars in a dense stellar environment, such as a nuclear or globular cluster.  However, the vast majority of merger rate estimates in the literature suggest that such dynamical formation is very rare relative to the event rate inferred from GW observations (see \citealt{MandelBroekgaarden:2021} for a review), and so is unlikely to be responsible for GW200115.   The only two exceptions are mergers in young star clusters \citep{Santoliquido:2020} (though other groups, e.g., \citealt{FragioneBanerjee:2020}, predict much lower merger rates) and mergers in hierarchical 3-body systems \citep{HamersThompson:2019}.  In the latter case, however, the BH spin is likely to be a consequence of binary evolution, discussed below, with the triple dynamics aiding the prompt merger and possibly contributing to spin misalignment if the BH is spinning \citep{LiuLai:2018,RodriguezAntonini:2018}.  Formation from first-generation population III stars or through chemically homogeneous evolution is similarly disfavored for BH -- NS binaries.  Therefore, we turn to the classical isolated binary evolution channel involving mass transfer.

The standard pathway for merging BH -- NS formation through isolated binary evolution proceeds as follows: $(i)$ binary formation, $(ii)$ mass transfer from the primary (initially more massive star) onto the secondary after the primary evolves off the main sequence, $(iii)$ collapse of the primary into a BH, $(iv)$ dynamically unstable mass transfer (a common-envelope phase) from the secondary onto the BH after the secondary evolves off the main sequence, $(v)$ possibly another phase of mass transfer from the stripped secondary after the end of the helium main sequence, $(vi)$ supernova explosion of the secondary leading to NS formation, and $(vii)$ a GW driven merger (see, e.g., Figure 3 of \citealt{Broekgaarden:2021} for an illustration).  In this process, the black hole's progenitor is stripped of its envelope, which contains the bulk of its moment of inertia and, hence, the bulk of angular momentum, assuming at least moderately efficient angular momentum transport, as supported by theory and observations \citep[e.g.,][]{Spruit:2002,FullerMa:2019,Belczynski:2020}.  The removal of the envelope leaves a naked helium star with little angular momentum, which is further reduced by spin-down through stellar winds.   At this stage, the helium star is too far away from the companion to be spun up through tidal interactions \citep{Kushnir:2016,Zaldarriaga:2017,HotokezakaPiran:2017,Qin:2018,Bavera:2019,Bavera:2020}.   Several possibilities for BH spin-up have been proposed, such as through supernova fallback torqued by the binary companion \citep{Schroeder:2018} or extreme super-Eddington accretion (requiring the BH mass to approximately double after formation), but most involve a degree of fine-tuning or assumptions that do not appear to be supported by current observations (see \citealt{MandelFragos:2020} for a critical summary).  It thus appears that this standard channel must inevitably yield a negligibly spinning BH \citep{BroekgaardenBerger:2021}.

There are a several variations that are worth considering.  It is possible that the NS forms first, before the BH.  This requires mass ratio reversal, so that the secondary forms a heavier remnant (the BH) than the primary (which form the NS).  In this case, the progenitor of the BH which forms from the secondary could be tidally spun up, since it would be in a close post-common-envelope binary with the NS, possibly leading to a rapidly spinning BH \citep{Debatri:2020}.  Population synthesis estimates by \citet{Broekgaarden:2021} suggest that mass ratio reversal is quite rare, comprising $\lesssim 1\%$ of all merging BH -- NS binaries.  The frequency of mass ratio reversal rises to $\approx 20\%$ under the assumption that Hertzsprung gap donors could initiate and survive common envelopes.  However, this assumption is disfavored by current understanding of mass transfer from stars without deep convective envelopes \citep{Klencki:2020convective}.  Furthermore, even if such a BH were rapidly spinning, it would be extremely unlikely to be misaligned from the binary's orbital angular momentum since it would form in a very tight binary after being aligned by tides and hence even a moderate supernova natal kick would not produce appreciable misalignment.  Meanwhile, a double-core common-envelope event could yield a tight binary with the possibility for tidally spinning up both cores, but this is not expected to produce significant numbers of BH -- NS binaries because this channel requires very similar companion masses; \citet{Broekgaarden:2021} estimate the contribution as $<1\%$ for all model variations.  If BHs that avoid mass transfer could have significant spins, another possible formation channel that could conceivably give rise to rapidly spinning BHs in BH -- NS binaries involves starting in a very wide binary and avoiding mass transfer altogether prior to the first supernova, relying on fortuitous natal kicks to bring the binary close.  This is, again, expected to be extremely rare.   

Perhaps the most promising scenario for forming a rapidly spinning BH in a BH -- NS system is one in which they arise from BH high-mass X-ray binaries.  These are systems comprising a black hole accreting winds from a massive stellar companion.  BHs in black-hole high-mass X-ray binaries Cygnus X-1, LMC X-1 and M33 X-7 appear to be very rapidly spinning (see, e.g., \citealt{MillerMiller:2015} for a discussion of the measurements and possible caveats).  A plausible scenario for the formation of such rapidly spinning BHs is that these were initially tight binaries with orbital periods of only a few days, so that the first stage of mass transfer from the primary began when the primary was still on the main sequence.  In that case, it may be possible for the donor, whose core and envelope are still tightly coupled, to simultaneously lose the bulk of its hydrogen envelope and get tidally spun up \citep{Valsecchi:2010,Qin:2019}.  Regardless of the formation mechanism, systems such as Cygnus X-1 could yield merging BH -- NS binaries in their future evolution \citep{Belczynski:2011CygX1}.  However, the donors in all observed BH high-mass X-ray binaries are nearly Roche-lobe filling (though this appears to be a consequence of the angular momentum content in accreted winds in such systems, \citealt{HiraiMandel:2021}).  Thus, these systems are on the verge of mass transfer while the secondary donor is still a main-sequence star.  Consequently, these systems appear unlikely to enter and survive a common-envelope phase, and may therefore remain too wide to merge through GWs.  For example, \citet{Neijssel:2020CygX1} estimate the future merger probability of Cygnus X-1 at only a few percent, which would rely on a favorable natal kick accompanying the second supernova.  On the other hand, such a natal kick could naturally explain misalignment along with the BH spin; \citet{Chia:2021} proposed this as a possible formation channel for the BH--BH merger GW151226, which may show evidence of both primary BH spin and misalignment (but see \citealt{GW151226,MateuLucena:2021}).  Simple estimates suggest that this channel could yield merger rates of a few Gpc$^{-3}$ yr$^{-1}$, which is 1--2 orders of magnitude lower than the BH -- NS merger rate inferred from GW200105 and GW200115 \citep{GW200105}.

We thus conclude that the presence of significant BH spin is not expected in the vast majority of merging BH -- NS binaries.  Only binaries in which the BH progenitor was simultaneously stripped and spun up during mass transfer late on the main sequence stage appear to be promising candidates for BH -- NS mergers with rapidly spinning and potentially misaligned BHs.  Given their anticipated low contribution to the total merger rate, we propose alternative spin priors on BH -- NS merger analysis that would comprise a mixture of negligible BH spin ($\sim 95\%$) and the standard LVK analysis broad priors on spin and misalignment angles ($\sim 5\%$). 

\begin{figure}
\centering
\includegraphics[width=\columnwidth]{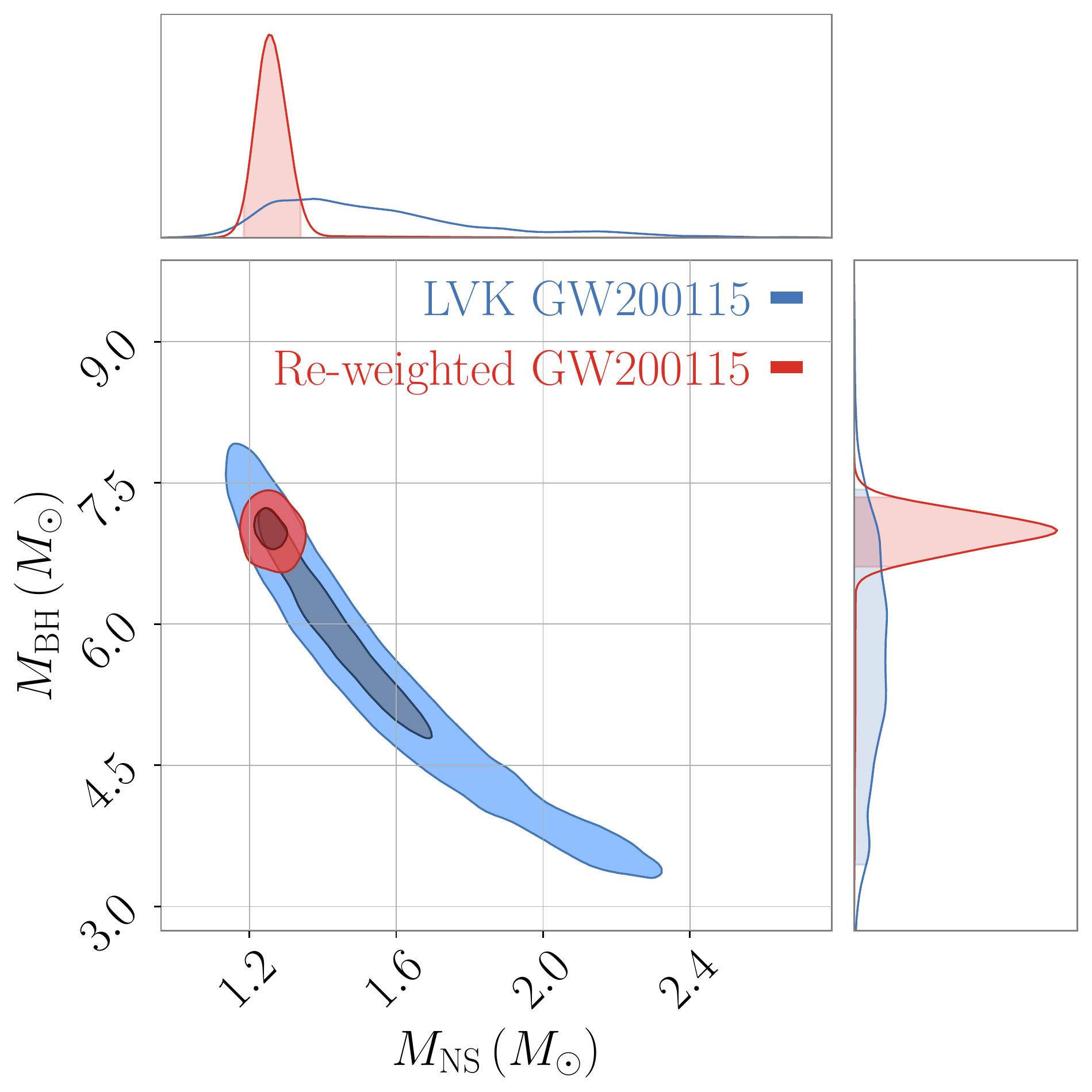}
\caption{Marginalized two- and one-dimensional posteriors in the space of $\MBH$ and $\MNS$.  
Blue shows the GW200115 posteriors as computed in \citep{GW200105} and red shows the GW200115 posteriors using the alternative priors proposed here.}\label{figure:m1m2}
\end{figure}

In practice, for the negligible-spin component, we treat the BH and NS spin magnitudes as being independent and identically distributed, following truncated narrow normal distributions peaked at zero. We do not modify the priors on spin angles. The prior distribution is 

\begin{equation}
\begin{aligned}
    \pi(\chi_\mathrm{BH}, \chi_\mathrm{NS}) =& 0.95\,\Big[\,\mathcal{N}(\chi_{\mathrm{BH}}; \mu=0, \bar{\sigma}=1.67\times10^{-2}) \times\\& \mathcal{N}(\chi_{\mathrm{NS}}; \mu=0, \bar{\sigma}=1.67\times10^{-2})\,\Big]\\ &+ 0.05\,\pi_{\text{LVK}}(\chi_\mathrm{BH}, \chi_\mathrm{NS})\,,
\end{aligned}
\end{equation}
where $\mathcal{N}$ is the normal distribution truncated to the interval $\big[0, 1\big]$, and with $\bar{\sigma}=1.67\times10^{-2}$ chosen so that each spin distribution has a standard deviation $\sigma_{\chi}=10^{-2}$. This choice is a practical one, to allow us to re-weight existing posterior samples: too few samples would be available for re-weighting for lower values of $\sigma_{\chi}$.  It is consistent with the negligible black hole spins observed in merging black holes that avoided tidal spin-up \citep{Galaudage:2021}.  These priors express the astrophysical a priori belief that the BH and NS are expected to have low spins.  At the same time, they allow us to re-weight the LVK analysis posterior samples.  Since there are a finite number of posterior samples with near-zero spins, and none with spin values of exactly zero, more narrow $\chi$ priors would create additional practical challenges.

We re-analyse the GW200115 signal with these astrophysically motivated priors by re-weighting the posterior samples from \citet{GW200105}.  We recover the $\chiz$ and $q$ posteriors shown in red in figure \ref{figure:qchiz}.  These are centered on zero spin: $\chiz=0.00^{+0.04}_{-0.04}$.  While this is not surprising given the prior preference for zero spin, the almost complete lack of a tail extending to either positive or negative $\chiz$ shows that there is insufficient likelihood preference at non-zero spin values to overcome a moderate prior re-weighting. 

The better constrained spin inferred with the priors advocated here yields a more precise mass ratio $q=0.18^{+0.03}_{-0.02}$, which in turn leads to the more precise mass measurements shown in figure \ref{figure:m1m2}.  We find that with our choice of priors, the component masses are tightly constrained, with $\MBH =7.0^{+0.4}_{-0.4}\,M_{\odot}$ and $M_{\text{NS}}=1.25^{+0.09}_{-0.07}\,M_{\odot}$.  The latter value is consistent with the typical masses of second-born neutron stars in Galactic double neutron star systems which are observed as radio pulsars.  In these Galactic systems, the second-born neutron stars are likely formed through ultra-stripped supernovae, with an extra episode of mass transfer (step $\it{v}$ in our standard pathway) from the NS progenitor after the end of its helium main sequence \citep{Tauris:2017}, possibly suggesting a similar formation channel in GW200115.

\acknowledgements
We thank Team COMPAS; Javier Roulet, Horng Sheng Chia, Matias Zaldarriaga and colleagues; Ben Gompertz and colleagues; Xingjiang Zhu; and Eric Thrane for useful discussions.  The authors acknowledge support from the Australian Research Council Centre of Excellence for Gravitational  Wave  Discovery  (OzGrav), through project number CE17010004. IM is a recipient of the Australian Research Council Future Fellowship FT190100574. This document has LIGO document ID P2100346.

\bibliographystyle{hapj}
\bibliography{Mandel}

\end{document}